\documentclass[12pt]{article}
\usepackage{epsf}
\title{ Mixing of meson,
hybrid, and glueball states }
\author{Yu.A.Simonov\\Institute of
Theoretical and Experimental Physics, \\ Moscow, Russia}
 \date{}
\newcommand{\be}{\begin{equation}}
 \newcommand{\ee}{\end{equation}}
 \def\la{\mathrel{\mathpalette\fun <}}

\def\fun#1#2{\lower3.6pt\vbox{\baselineskip0pt\lineskip.9pt
\ialign{$\mathsurround=0pt#1\hfil
##\hfil$\crcr#2\crcr\sim\crcr}}}
\newcommand{\veX}{\mbox{\boldmath${\rm X}$}}
\newcommand{\veY}{\mbox{\boldmath${\rm Y}$}}
\newcommand{\vex}{\mbox{\boldmath${\rm x}$}}
\newcommand{\vey}{\mbox{\boldmath${\rm y}$}}
\newcommand{\ver}{\mbox{\boldmath${\rm r}$}}

\newcommand{\veP}{\mbox{\boldmath${\rm P}$}}
\newcommand{\vep}{\mbox{\boldmath${\rm p}$}}

\newcommand{\vez}{\mbox{\boldmath${\rm z}$}}
\newcommand{\veS}{\mbox{\boldmath${\rm S}$}}
\newcommand{\veL}{\mbox{\boldmath${\rm L}$}}
\newcommand{\veR}{\mbox{\boldmath${\rm R}$}}

\newcommand{\vek}{\mbox{\boldmath${\rm k}$}}
\newcommand{\ven}{\mbox{\boldmath${\rm n}$}}

\newcommand{\vexi}{\mbox{\boldmath${\rm \xi}$}}
\newcommand{\veta}{\mbox{\boldmath${\rm \eta}$}}
\newcommand{\veB}{\mbox{\boldmath${\rm B}$}}

\newcommand{\veE}{\mbox{\boldmath${\rm E}$}}

\newcommand{\vegam}{\mbox{\boldmath${\rm \gamma}$}}

\newcommand{\lan}{\langle}
\newcommand{\ran}{\rangle}
\begin{document}
\maketitle
\large
{\small The effective QCD Hamiltonian is constructed with the help
of the background perturbation theory, and relativistic
Feynman--Schwinger path integrals for Green's functions. The
resulting spectrum displays mass gaps of the order of  one GeV,
when additional valence gluon is added to the bound state.}

{\small Mixing between meson, hybrid, and glueball states is
defined in two ways: through generalized Green's functions and via
modified Feynman diagram technic giving similar answers. Results
for mixing matrix elements are numerically not large (around 0.1
GeV) and agree with earlier analytic estimates and lattice
simulations.}


\section{Introduction}

Gluonic fields play a double role in the dynamics of QCD. On one
hand they create the QCD string between color charges which
substantiates confinement. On another hand valence gluons act as
color sources and can be considered as constituent pieces of
hadrons on the same grounds as quarks. This is a QCD string
picture of hybrids and glueballs developed analytically in
\cite{1} and confirmed by recent lattice calculations \cite{2}. On
experimental side the study of hybrid \cite{3} and glueball
\cite{4} states is not yet conclusive, and more  experimental and
theoretical work is needed, especially a quantitative treatment of
mixing between mesons, hybrids, and glueballs.

One typical feature of the QCD string approach, supported by
lattice data \cite{2}, is that the addition of every valence gluon
in a hadron increases the hadron mass by 0.8 -- 1 GeV, and this is
true also for purely gluonic states -- glueballs. Thus a large
mass gap exists between meson ground state and its gluonic
excitation, which makes it possible to consider gluonic admixture
in a given hadron state as a perturbation. Therefore in the zeroth
approximation one has a Hamiltonian for the diagonal states of the
fixed number of quarks $q, \bar q$ and valence gluons $g$, while
in the next approximation one calculates the  mixing between the
states perturbatively (unless masses of the states  happen to be
almost degenerate, in which case one solves a matrix Hamiltonian).

The mixing between meson and hybrid states was considered
previously analytically in \cite{5,6}. In \cite{5} the hybrid wave
function was taken in the cluster approximation, with gluon wave
function of bag-model type factorized using  the $q\bar q$
potential model wave function.  The resulting mixing matrix
elements (MME) are not large, supporting the iterative scheme
described above. A similar approach, based on the nonrelativistic
constituent quark and gluon model, was used in \cite{6} to study
mixing between $1^- c\bar c g$ hybrid meson and charmonium.

The mixing between glueballs and other hadrons was studied on the
lattice \cite{7}, with  reasonably moderate MME.

It is a purpose of the present paper to develop a general
formalism of the QCD bound states including valence gluons, with
the special attention to the mixing between states. This formalism
is based on the QCD string approach \cite{1} and applicable to
both relativistic and nonrelativistic systems, massive or massless
quarks and gluons\footnote{A short version of an approach in the
same direction with a different form of matrix elements appeared
recently \cite{8}. Results of \cite{8} are similar to ours
qualitatively, with some numerical differences.}.

Two features are important for this formalism. First, it is
derived directly from QCD with few assumptions, supported and
checked by lattice computations. Second, the only input of the
theory is the fundamental string tension $\sigma_f$, current quark
masses (renormalized at the scale of 1 GeV), and strong coupling
constant $\alpha_s$.  (For the total hadron masses one needs to
subtract a selfenergy for each quark and antiquark equal to 0.25
GeV, but this does not affect wave functions and mixings which
will be the main goal of this paper).

The plan of the paper is as follows. In the next section we
develop the background perturbation theory to separate valence
gluons from the background gluonic fields forming the QCD strings.
On this basis the diagonal part of  the QCD Hamiltonian is written
for quarks, antiquarks,  and valence gluons interacting via string
connection and main characteristics of the spectrum are
established for mesons, hybrids, and glueballs.

In section 3 the part of the Hamiltonian responsible for the
mixing of mesons, hybrids, and glueballs is identified and MME is
written in terms of solutions of the diagonal Hamiltonian, using
formalism of modified plane  waves.

In section 4 concrete calculations of MME are presented for
meson-hybrid case based on the Green's function formalism. An
analogous treatment of the meson-glueball case is given in section 5.

In concluding section discussion is given of the results obtained
in this paper and comparison is made with  calculations with ohter
model, lattice simulations, and experimental data.

\section{ Green's functions and Hamiltonian}

We follow in this section the procedure developed in detail in
\cite{9,10} ( for earlier refs. see \cite{11}) and therefore here
we recapitulate only the main points. The total gluonic field
$A_\mu$ is represented as a sum \be A_\mu= B_\mu+a_\mu, \label{1}
\ee where $B_\mu$ is the nonperturbative part and $a_\mu $ --
perturbative (or valence gluon) part, which shall be treated in
form of the perturbation  series  in $ga_\mu$. One can formally
avoid the problem of doublecounting using the 't Hooft identity,
\cite{9} which allows to represent the partition function  as \be
Z=\frac{1}{N}\int e^{-S(A)} Dq D\bar q DA= \frac{1}{N'}\int \eta
(B) e^{-S(B+a)} DB Da Dq D\bar q, \label{2} \ee where $N,N'$ are
normalization constants.

 Here $S$ is total Euclidean action and
$\eta (B)$ -- an arbitrary weight of averaging over $DB$.
Expanding $S(B+a)$ one obtains \be S(B+a) = S(B) + S_1(a, B)+
S_2(a, B) + S_3(a, B)+S_4(a) + S_1(a,q,\bar q) \label{3} \ee
 In (\ref{3}) $S_n(a,B)$ denotes terms of power $n$
in $a_\mu$, and $S_1(a,q, \bar q)$ is the mixing term, which will
be of main interest in  section 3.  The terms $S_3$ and $S_4$
contain additional powers of $g$ and can be treated
perturbatively, the term $S_1(a,B)$ was considered in \cite{10}
and shown  to yield a small   correction to the  leading terms
$S(B)$ and $S_2(a,B)$, the latter defining the valence gluon
propagating in the background $B_\mu$ with the Green's function
(in the background Feynman gauge \cite{9}--\cite{11}) \be
G^{(g)}_{\mu\nu}= (D^2_\lambda\delta_{\mu\nu}+ 2ig
F_{\mu\nu})^{-1},~~D_\lambda =\partial_\lambda-ig B_\lambda,
F_{\mu\nu}^a =\partial_\mu A_\nu^a-\partial_\nu A_\mu^a+ gf^{abc}
A_\mu^bA^c_\nu. \label{4}
 \ee
  The explicit form of all terms in (\ref{3}) and of
the corresponding Green's functions is given in \cite{9,10}, here
we only quote the results for the Green's function of the state
containing quark, antiquark, and any number of valence gluons in
the  leading $N_c$ approximation: \be
 G_{q\bar q(ng)} (X,Y) \sim
\lan\Gamma_{in} S_q (x,y) \prod^n_{i=1} G^{(g)}_{\mu_i\nu_i} (x,y)
S_{\bar q} (x,y) \Gamma_f\ran_B.
 \label{5}
  \ee
  For
(multi)glueball Green's function one has a similar representation
with missing factors $S_q, S_{\bar q}$. Here $\Gamma_{in},
\Gamma_f$ are initial and final vertex operators and $\lan...
\ran_B$ means averaging over fields $B_\mu$ with the weight
$\eta(B)$, and quark Green's function $S_q$ can be written as \be
S_q= -i(\hat D+m)^{-1} = -i(m-\hat D)(m^2-\hat D^2)^{-1}=
-i(m-\hat D)(m^2-\hat D^2_\mu -\sigma_{\mu\nu} F_{\mu\nu})^{-1}.
\label{6} \ee In what follows we shall treat spin-dependent terms
for gluon and quark Green's functions in (\ref{4}) and (\ref{6})
as a small correction, which is supported by exact calculations
and comparison with experiment and lattice data \cite{1}.
Correspondingly one can use for the quadratic in $D_\mu$ parts the
Feynman--Schwinger (world-line) path-integral representation (see
\cite{12,13} and refs. therein) \be
(m^2-D^2_\mu)_{xy}^{-1}=\int^\infty_0 ds \int (Dz)_{xy} e^{-K}
\exp (ig \int^x_y B_\mu dz_\mu), ~~ K=\frac14 \int^s_0
(\frac{dz_\mu}{d\tau})^2 d\tau. \label{7} \ee It is important that
background field $B_\mu$ enters (\ref{7}) only in the exponentials
and one can show \cite{12} that in the total Green's function
(\ref{5}) in the leading $N_c$ approximation all these
exponentials combine into products of Wilson loops. Here we make
another assumption, supported by lattice data, that Wilson loop
has a minimal area law behaviour with the string tension
$\sigma_f$ (for mesons and hybrids) and $\sigma_{\rm{adj}}
=\frac94 \sigma_f$ for $2g$ glueballs. Thus one reduces the system
of $q,\bar q$ and $n$ gluons to the open string with the ends at
$q$ and $\bar q$, and $n$ gluons ``sitting" on the string, and
glueballs reduce  to an open adjoint string (or equivalently for
large $N_c$ -- a closed fundamental string).

At this point one can define the Hamiltonian of the system $H$,
using the relation for the general Green's function of the type
(\ref{5}) \be G(X,Y)=\lan X|\exp (-HT)|Y\ran
\label{8}
 \ee
 where
$T=X_4-Y_4$.

The Hamiltonian $H$ can be defined on any hyperplane, in case of
center-of-mass system one gets the Hamiltonian which was obtained
for the $q\bar q$ system in \cite{14} and generalized in \cite{17}
to hybrid and in \cite{16} to the glueball case (see also
\cite{15} for earlier papers on hybrid Hamiltonian).

Here we quote the simplest version of the Hamiltonian for $q\bar
q$ and $n$ gluons, where string rotation contribution to the
moment of inertia is treated as a perturbation: \be H=H_0+\Delta
H_L + \Delta H_s+ \Delta H_c \label{9} \ee where we have defined.
\be H_0=\frac{\mu_q+\mu_{\bar q}}{2}+\sum^n_{k=1}\frac{\mu_{k}}{2}
+\sum_{k=q,\bar q,1...n} \frac{\vep^2_k+m^2_{k+1}}{2\mu_k}+
\sigma_f\{ |\ver_q-\ver_1|+ |\ver_1-\ver_2|+
...+|\ver_n-\ver_{\bar q}|\}.
 \label{10}
 \ee
  Here
$m_k =0 $ for $ k=1,..., n$ and $\mu_k$ defined as
$\frac{1dz_4}{2d\tau_k}$ (see [14]) are to be found from the
minimum of the $H_0$, which is approximated within 5\% accuracy by
the minimum of the eigenvalue $M_0^{(\nu)}$ \cite{18},
$H_0\Psi_\nu=M_0^{(\nu)}\Psi_\nu$:
 \be \frac{\partial
M_0^{(\nu)}}{\partial\mu_k}\left\vert_{\mu_k=\mu_k^{(0)}}\right.=0,~~
k=q,\bar q, 1,...,n; \nu=0,1,2,... \label{11} \ee $\Delta H_L$
gives zero contribution when all interparticle angular momenta are
zero, and in particular for the meson case has the form \cite{14}:
\be \Delta H_L=-\frac{16\sigma^2 L(L+1)}{3 M^3_0}. \label{12} \ee
The Coulombic part $\Delta H_c$ takes into account lowest-order
Coulomb exchanges between quark and antiquark. It is argued in
\cite{16} that Coulomb exchanges between valence gluons are
strongly suppressed by higher-loop corrections and hence can be
neglected in the first approximation.

Finally, the spin-dependent term $\Delta H_s$ has the following
form \cite{1} for $q\bar q$ or $gg$ case \be \Delta H_s=
a\sum_{i,k=1,2}\frac{\veS_i\cdot \veS_k}{\mu_i\mu_k}+
b_1\sum_{i}\frac{\veS_i\cdot \veL_i}{\mu_i^2}+
\frac{b_2}{\mu_1\mu_2}\sum_{i}\veS_i\cdot \veL_i+
c\frac{3(\veS_1\cdot \ven)(\veS_2\cdot \ven)-\veS_1\cdot
\veS_2}{\mu_1\mu_2}. \label{13} \ee

In the c.m.s. $\veL_1=-\veL_2=\veL$ $a,b_1, b_2$, and $c$ depend
on distance $r$ between $q\bar q (gg)$ and can be found in
\cite{1}.

Let us now discuss the properties of the resulting Hamiltonian
(\ref{9}). First of all, one should stress that it is a fully
relativistic Hamiltonian. Indeed, neglecting for the moment the
interaction term in (\ref{10}) and finding $\mu_k$ from (\ref{11})
one immediately obtains $\mu_k=\sqrt{m^2_k+\vep^2}$, i.e., in the
free case $\mu_k$ plays the role of the relativistic energy of a
quark or gluon. Moreover, the form of $\Delta H_s$ is not a result
of expansion in inverse powers of large masses $\mu_i, \mu_k$, but
rather is a result of Gaussian approximation for the average of
spin-dependent factors $\lan\prod_n\exp
(g\sigma^{(n)}_{ik}F^{(n)}_{ik})\ran_B$ (see \cite{1} for
details).

It is important to stress that $m_q, m_{\bar q}$ entering the
Hamiltonian (\ref{10}) denote quark current masses, renormalized
at the scale around 1GeV, and  nowhere we use as input constituent
masses of quarks or gluons.

The spectrum of the Hamiltonian (\ref{9}) was calculated for many
systems, including mesons, hybrids, and glueballs, for a review
see \cite{1}. As a recent example see calculation of gluelumps in
\cite{19}.

The characteristic feature of the spectrum is that each
constituent (quark or gluon) contributes to the total mass a
quantity approximately equal to $2\mu^{(0)}_k$, where
$\mu^{(0)}_k$ depends on the number of eigenvalue $\nu$ and is
expressed through $\sigma_f$ (or $\sigma_{\rm adj}$ for
glueballs). $\mu_k^{(0)}$ changes from 0.35 GeV for massless
quarks in the lowest meson states, and $\sqrt{\frac94}0.35=0.52$
GeV for gluons in the glueball. Therefore any gluon in a hybrid
adds around 1 GeV to the total mass, and the same is true for
glueballs: the lowest three-gluon glueball is approximately 1 GeV
heavier than  the lowest two-gluon glueball. This fact supports
the idea outlined in the Introduction that the diagonal
eigenvalues of the total Hamiltonian have energy gaps around 1 GeV
and it may be a good approximation to treat mixing due to valence
gluon excitation as a perturbation provided the MME is much less
than 1 GeV.

\section{Mixing matrix elements}

We turn in this section to the part of interaction, $S_1(a,q,\bar
q)$, which is responsible for the mixing between hadronic states,
differing by the number of valence gluons. It has the form \be
S_1(a,q,\bar q)= g\int\bar q(x) \hat a(x) q (x) d^4 x. \label{14}
\ee
 Here $\hat a(x)=a^a_\mu t^a_{\alpha\beta}\gamma_\mu$, and
$\bar q_\alpha, q_\beta$ all have color indices, whereas in the
Hamiltonian (\ref{10}) and its eigenfunctions the color indices
are absent because of color averaging in $G_{q\bar q (ng)}$
(\ref{5}) and hence in $H$.

To understand how the matrix element of $S_1$ is taken between
colorless hadronic eigenfunctions of $H$, one can use the
formalism of Green's functions, described in Sections 4,5, and
corresponding to the diagrams in Figs. 1--4.

Here we shall choose a simpler way, which leads to the same
results as in sections 4,5, but more familiar for the reader
accustomed to Feynman diagrams.

The rules are simple: represent each $q(x), \bar q(x)$ and
$a_\mu(x)$ by equivalent colorless fields with familiar plane-wave
expansion \be q(\vex, t) =\sum_{\vek}\frac{1}
{\sqrt{2\mu_q(\vek)V}} [(u(k,\sigma) b_\lambda \exp({i\vek\cdot
\vex-i\mu_q t})+u^c(k,\sigma) d_\lambda^+\exp({-i\vek\cdot
\vex+i\mu_q t}) )] \label{15}
 \ee
 \be
 a_\mu(\vex, t)
=\sum_{\vek,\lambda}\frac{1}{\sqrt{2\mu(\vek)V}} [\exp({i\vek\cdot
\vex-i\mu t)}e^{(\lambda)}_\mu c_\lambda (\vek)+e^{(\lambda)}_\mu
c^+_\lambda(\vek) \exp({-i\vek\cdot \vex+i\mu t})],
 \label{16}
 \ee
 where $b_\lambda$ and $c_\lambda$ are annihilation
operators for the quark and gluon respectively, $d^+_\lambda$ is
creation operator for antiquark and $u(k)$ is normalized according
to the condition $\bar u(k) u(k)=2\mu_q(k),$ $V$ - is the 3d
volume and $e^{(\lambda)}_\mu$ is the gluon polarization vector.

One can easily check at this point that introduction of (\ref{16})
into the free part of the QCD Hamiltonian
$\frac{\veE^2+\veB^2}{2}$ immediately reproduces the gluonic part
of the effective Hamiltonian (\ref{10}) (the same is true for
$q,\bar q$ if one uses the quadratic in $\vek$ form of the
Hamiltonian).

In (\ref{15}), (\ref{16}) $\mu_q(\vek)$
 and $\mu(\vek)$  are the same quantities as in
(\ref{10}), to be defined by the minimization procedure in the
system, where quark or gluon enter as constituents.

Another important point is the number of independent polarizations
of a bound gluon. We shall use as in \cite{9}--\cite{11} "the
gauge-invariant gauge condition" \be D_\mu a_\mu=0. \label{17} \ee

For a valence gluon in the gauge-invariant hybrid wave function
$\Psi_\mu$ one has \be \Psi_\mu(x,z,y)\equiv \bar q(x) \Phi(x,z)
a_\mu(z) \Phi(z,y) q(y). \label{18} \ee Here $\Phi$ are parallel
transporters depending on $B_\mu$, $\Phi(x,y)=P\exp ig \int^x_y
B_\mu dz_\mu$.

Differentiating $\Psi_\mu$ in $\frac{\partial}{\partial z_\mu}$
one obtains additional term $B_\mu$ due to differentiation of the
end points in $\Phi(x,z), \Phi(z,y)$, so that one has \be
\frac{\partial}{\partial z_\mu} \Psi_\mu(x,z,y) = \bar q(x)
\Phi(x,z) (D_\mu a_\mu(z))\Phi(z,y) q(y)+... \label{19} \ee where
dots imply the terms from the contour differentiation, i.e., due
to additional gluonic excitation of the string.

In the diagonal approximation (and keeping in mind that gluonic
excitation implies a gap of 1 GeV), we disregard those terms, and
hence have the  (approximate) condition \be
\frac{\partial}{\partial z_\mu}\Psi^{(h)}_\mu(x,z,y) =0.
\label{20} \ee This means that only 3 gluon polarizations are
physical, i.e., the same situation as for  an off-shell photon or
gluon. In what follows we shall retain only\\ $\mu=1,2,3$. It is
clear that the same reasoning applies to gluon in glueball.

We are now in the position to write the general form of wave
functions for mesons, hybrids, and glueballs. In the momentum
space and in the second-quantized form they can be written for a
meson \be \phi^{(M)} =f^{(\sigma_1,
\sigma_2)}_{\alpha_1,\alpha_2}(\vep_1,\vep_2)b^+_{\sigma_1}(p_1)
d_{\sigma_2} (p_2), \label{21} \ee where $\sigma_1,\sigma_2$ are
polarizations (helicities) of $q$ and $\bar q$, and $\alpha_1,
\alpha_2$ are Dirac 4-spinor indices.

For a hybrid one has
\be
\phi^{(H)}
=f^{(\sigma_1, \sigma_2,
\lambda)}_{\alpha_1,\alpha_2,\mu}(\vep_1,\vep_2,\vep_3)
b^+_{\sigma_1}(p_1) d_{\sigma_2} (p_2) c^+_\lambda(p_3)
 \label{22}
 \ee
where $\lambda$ is the gluon polarization and
$\mu$ is discussed above, it enters as in
$e^{(\lambda)}_\mu$ in (\ref{16}). Finally for the
glueball one has
\be
\phi^{(G)}
=f^{(\lambda_1,
\lambda_2,)}_{\mu_1,\mu_2}(\vep_1,\vep_2)
c^+_{\lambda_1}(p_1)
c^+_{\lambda_2}(p_2).
 \label{23}
 \ee
One should note, that all wave functions (\ref{21})--(\ref{23})
are given in the representation when the total angular momentum
$J$ and its projection $M_J$ are not projected out. In fact the
operator (\ref{21}) is a 16 component structure, and one can use
the classification introduced in \cite{21}  to distinguish
positive and negative energy states of quarks using the so-called
$\rho$-spin, $\rho=\pm$ and usual spin states for each quark. From
spin and $\rho$-spin states one can construct the states with
given  total momentun and parity \cite{22}. These scheme was
exploited and  developed  in the series of papers \cite{23} where
the $q\bar q$ interaction was used containing both scalar and
vector confining parts.

A similar scheme can be used for hybrids, but the resulting
calculations are rather cumbersome. Therefore in this section we
only list some schematic expressions with coefficients for mesons
which can be found in \cite{23} and for hybrids not yet available
(to the knowledge of the author). Thus one can write instead of
wave functions (\ref{21})--(\ref{23})  the functions
$\Phi^{(M)}_J, \Phi^{(H)}_J$, and $\Phi^{(G)}_J$ with given total
angular momentum. Each of this functions consists of several
components differing in total spin, orbital momentum, and
$\rho$-spins. It is actually those combinations which should be
inserted everywhere below in this chapter instead of $\Phi^{(M)},
\Phi^{(H)}$, and $\Phi^{(G)}$, respectively, but we keep for
simplicity reason the functions (\ref{21})--(\ref{23}), referring
the reader to the Appendix for more details and discussion.

We are now in position to write down the MME between states
(\ref{21})--(\ref{23}), using the interaction Hamiltonian,
obtained from (\ref{14}), namely
$$
H_1=g\int
\bar q(\vex,0)\hat a(\vex, 0) q(\vex, 0) d^3x=
$$
\be =g\sum_{\vek_1,\vek_2, \vek_3} \bar q(\vek_1)\hat a(\vek_3)
q(\vek_2) \frac{ d^3k_1}{(2\pi)^3} \frac{ d^3k_2}{(2\pi)^3} \frac{
d^3k_3}{(2\pi)^3} {(2\pi)^3} \delta^{(3)}(\vek_1+\vek_2+\vek_3).
\label{24} \ee

For the meson-hybrid mixing one has
$$
(\Phi^{(M)}, H_1\Phi^{(H)})=g\int
f^{+(\sigma_1,\sigma_2)}_{\alpha_1,\alpha_2}(\vep_1,
\vep_2)\frac{\hat e^{(\lambda)}(k_3)}{\sqrt{2\mu(k_3)}}
f^{\sigma'_1,\sigma'_2,
\lambda}_{\alpha'_1,\alpha_2,\mu}(\vep'_1,\vep_2,\vek_3)\times
$$
\be \times \frac{ d\vep_1}{(2\pi)^3} \frac{ d\vep'_1}{(2\pi)^3}
\frac{ d\vek_3}{(2\pi)^3} \delta^{(3)}(-\vep_1+\vep'_1+\vek_3)
(2\pi)^3\frac{d^3p_2}{(2\pi)^3}(2\pi)^3\delta^{(3)}(\vep_1+\vep_2).
\label{25} \ee The normalization condition for $\Phi^{(M)},
\Phi^{(H)}$ looks like \be (\Phi^{(M)}, \Phi^{(M)})=\int
f^{+(\sigma_1,\sigma_2)}_{\alpha_1,\alpha_2}(\vep_1,
 \vep_2)
f^{(\sigma'_1,\sigma'_2)}_{\alpha_1,\alpha_2}(\vep_1, \vep_2)
dP_2= \delta_{\sigma_1\sigma'_1} \delta_{\sigma_2\sigma'_2},
\label{26}
 \ee
  \be
  (\Phi^{(H)}, \Phi^{(H)})=\int
f^{+(\sigma_1,\sigma_2,\lambda)}_{\alpha_1,\alpha_2,\mu}(\vep_1,
\vep_2,\vep_3)
f^{(\sigma'_1,\sigma'_2,\lambda')}_{\alpha_1,\alpha_2,\mu}(\vep_1,
\vep_2,\vep_3) dP_3=
\delta_{\sigma_1\sigma'_1}\delta_{\sigma_2\sigma'_2}
\delta_{\lambda\lambda'}, \label{27} \ee where we have defined \be
dP_3\equiv \prod^3_{i=1}
\frac{d^3\vep_i}{(2\pi)^3}(2\pi)^3\delta^{(3)}
(\vep_1+\vep_2+\vep_3), \label{28} \ee \be dP_2\equiv
\prod^2_{i=1} \frac{d^3\vep_i}{(2\pi)^3}(2\pi)^3\delta^{(3)}
(\vep_1+\vep_2). \label{29} \ee Going from momentum to coordinate
representation and suppressing for the moment the momentum
dependence of the factor $\Gamma (p_1, p'_1, k_3)$ in (\ref{25}),
where \be \Gamma (p_1, p'_1, k_3)\equiv \frac{\hat
e^{(\lambda)}(k_3)}{\sqrt{2\mu(k_3)}  }, \label{30} \ee one has
\be (\Phi^{(M)}, H_1\Phi^{(H)})=g\int
d^3\ver\varphi^{(M)}(\ver)\Gamma\varphi^{(H)}(0,\ver) \label{31}
\ee where we have defined
$$
\varphi^{(M)}(\ver)\equiv \int
f^{(\sigma_1,\sigma_2)}_{\alpha_1,\alpha_2}(\vep_1,\vep_2)
e^{i\vep_1\cdot \ver_1+i\vep_2\cdot \ver_2}dP_2=
$$
\be =\int
f^{(\sigma_1,\sigma_2)}_{\alpha_1,\alpha_2}(\vep_1,-\vep_1)
e^{i\vep_1\cdot \ver_1}
\frac{d\vep_1}{(2\pi)^2},~~\ver=\ver_1-\ver_2; \label{32} \ee \be
\varphi^{(H)}(0,\ver)= \int
f^{(\sigma_1,\sigma_2,\lambda)}_{\alpha_1,\alpha_2,\mu}(\vep_1,\vep_2,\vep_3)
e^{i\vep_1 \cdot \ver_1+i\vep_2 \cdot \ver_2+i\vep_3 \cdot
\ver_3}dP_3 \label{33} \ee with $ \ver_1-\ver_3=0,~~
\ver_1-\ver_2\equiv \ver.$

In the next two sections we shall use another formalism to derive
MME -- the formalism of Green's functions, which enables us to use
our Hamiltonian technic described in chapter 2.

   \section{
   Mixing between meson and hybrid states}

\begin{figure}[!t] 
\epsfxsize=12cm 
\centering
\epsfbox{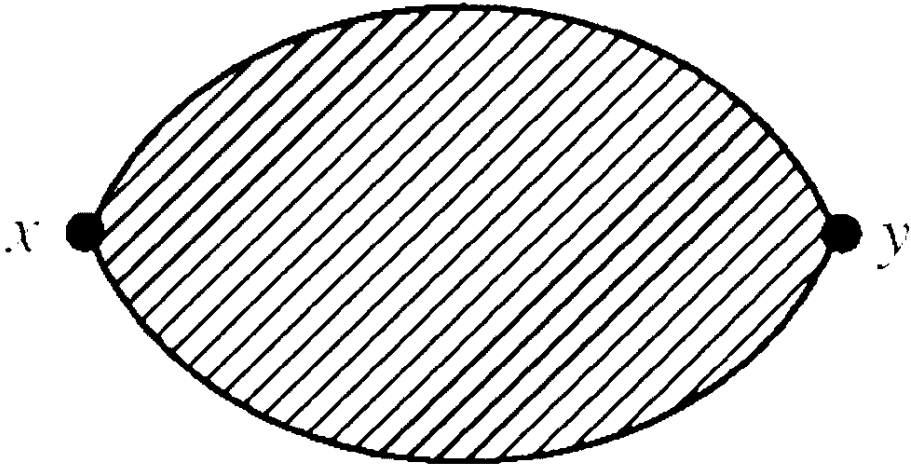} 
\caption{Meson Green's function describing  propagation of the quark and antiquark (solid
lines) from the point $x$ to the point $y$.
 The hatched interior of the figure
  implies presence of nonperturbative fields $B_\mu$ in the form of the
    fundamental string world sheet.}
\end{figure} 

   Consider a nonsinglet $q\bar q$ state; the corresponding Green's
   function in the quenched (large $N_c$) approximation can be
   written, according to general formulas \cite{12,13}, as
(see Fig. 1 for the corresponding Feynman diagram)
   $$
   G_{q\bar q} (x,y) = \lan
   \bar \psi(x) \Gamma^{({\rm out})} \psi(x)
   \bar \psi(y) \Gamma^{({\rm in})} \psi(y)\ran_{B,a,\psi,\bar\psi}=
     $$
     \be
     =\lan tr \Gamma^{({\rm out})} S(x,y) \Gamma^{({\rm in})} S(y,x)\ran_{B,a}.
     \label{A1.1}
     \ee
     Here averaging over $B,a$ is assumed with the action given in
     (\ref{2}),(\ref{3}) and $S(x,y)$ is the quark Green's function,
     \be
     S(x,y)= -i(\hat D+m)^{-1}_{x,y}, D_\mu=\partial_\mu-igB_\mu.
     \label{A1.2}
     \ee
     Writing $S=-i(m-\hat D)(m^2-\hat D^2)^{-1}$, one can reduce the
     Wilson-loop path integral for $G_{q\bar q} $ to the form
     \be
     G_{q\bar q} (x,y) = \lan tr_{c\alpha} \Gamma^{({\rm out})}(m-\hat D)
     \int^\infty_0
     \int^\infty_0
     ds d\bar s   e^{-K-\bar K} (Dz)_{xy} (D\bar z)_{xy}
     \Gamma^{({\rm in})} (\bar m-\hat{\bar D})W_F\ran .
     \label{A1.3}
     \ee
     Here $W_F$ is the Wilson loop with insertion of operators
     $\sigma F$, defined in \cite{6}, and trace is over color (c) and
     Lorentz $(\alpha)$ indices.

     Our primary task now is to transform the integral $\int ds
     (Dz)_{xy}$ of each quark Green's function as follows:
     \be
     ds(Dz)_{xy}= ds \prod^N_{n=1}\frac{d^4z(n)}{(4\pi\varepsilon)^2}
     \frac{d^4p}{(2\pi)^4} e^{ip(x-y-\sum_n z(n))},~~ N\varepsilon
     =s.
       \label{A1.4} \ee

     Introducing now the parameter $\mu(z_4)$ playing the role of
constituent quark mass according to
$$
ds=\frac{dz_4}{2\mu(z_4)},~~ \int ds\frac{dp_4}{2\pi}
e^{ip_4(\sum_n\Delta z_4^{(n)}-T)}= \int ds \delta(2\bar \mu s -T)
=\frac{1}{2\bar \mu},
$$
  where $2\bar\mu s=\sum_n\Delta z_4(n) =
  \sum_n\Delta s 2\mu(n)$,  one obtains
  \be
  Ds(Dz)_{xy}=\frac{1}{2\bar\mu}\frac{d^3p}{(2\pi)^3}
  e^{i\vep\cdot (\vex-\vey-\sum_n\vez(n))} D^3z D\mu.
  \label{A1.6}
  \ee
  Thus the path integral for each Green's function acquires the 3-d
  form, which we call $G_{(3d)}$, with  all Zitterbewegung ($z$-graphs) contained in the integral over $D\mu$.  If one does, as we
  usually do for all systems (except pion and kaon), the stationary
  point procedure in integration over $D\mu$, then one finds the
  smooth "constituent mass trajectory" $\mu_0(t)(t\equiv z_4)$, or a
  simple approximation to it, the constant constituent mass $\bar
  \mu_0$, found from the minimum of the Hamiltonian eigenvalue
  \cite{20}. In this approximation (yielding accuracy around 5\% for
  masses \cite{18}) one can identify $\bar \mu$ in (\ref{A1.6}) and
  $\bar \mu_0$.

  Thus one can write
  \be
  S(x,y) = \frac{m-\hat D}{2\bar \mu_q} G_{3d} (x,y) = \frac{m-\hat
  D}{2\bar \mu_q} \lan x|n\ran e^{-M_n|x-y|}\lan n| y\ran,
  \label{A1.7}
  \ee
  where $M_n$ is the n-th eigenvalue of the Hamiltonian.
  Similarly for the gluon Green's function one writes
  \be
  G_{\mu\nu} (x,y) = (-\hat D^2\delta_{\mu\nu}-2i\hat
  F_{\mu\nu})^{-1}_{xy} =
  \frac{1}{2\bar \mu_g}
   \lan x,\mu |n\ran e^{-M_n|x-y|}\lan  n| y, \nu\ran.
  \label{A1.8} \ee

  The form in (\ref{A1.7}),(\ref{A1.8}) is highly symbolic, since
  quarks and gluons do not propagate separately (and do not have
  separate eigenfunctions $(|n\ran) $, but rather form the common
  bound state. In the free case  one can identify $\bar \mu_q$ with
  the energy $E_q, \bar \mu_g$ with $\omega_g$, and
  $|n\ran =\frac{\exp({i\vep\cdot \vex})}{(2\pi)^{3/2}}$, so that
  (\ref{A1.7}), (\ref{A1.8}) go over into well-known representation
  \be
  S(x,y) =-i (m-\hat \partial)
  \int\frac{d^3p}{(2\pi)^3}\frac{\exp({i\vep
  \cdot \vex-E|x_4-y_4|)}}{2E(p)},
  \label{A1.9}
  \ee
  \be
  G_{\mu\nu} (x,y)=\delta_{\mu\nu}\int \frac{d^3\vek
  \exp({i(\vek\cdot \vex-\omega|x_4-y_4|)})}{(2\pi)^32\omega(k)}.
  \label{A1.10}
  \ee
  Consider now the hybrid Green's function
  \be
G^{(h)} (x^{(1)}, x^{(2)}, x^{(3)}| (y^{(1)}, y^{(2)}, y^{(3)}) =
\lan \Psi^{(h)+} (x^{(1)}, x^{(2)}, x^{(3)}) \Psi^{(h)}(y^{(1)},
y^{(2)}, y^{(3)})\ran_{B,a,q}.
\label{A1.11}
 \ee
 Omitting parallel
transporters for simplicity, one can write \be
G^{(h)}=\sum^\infty_{n=0}\lan\vex^{(1)},\vex^{(2)},
\vex^{(3)}|n\ran \Lambda_q^{(1)}\Lambda^{(2)}_{\bar
q}\frac{e^{-M_nT}}{2\bar \mu_g(n)} \lan n|\vey^{(1)},
\vey^{(2)},\vey^{(3)}\ran, \label{A1.12} \ee where we have denoted
$\Lambda_q^{(1,2)}=\frac{m-\hat D}{2\bar \mu_q}$ for quark and
antiquark respectively.

The Hamiltonian eigenfunctions $ \lan n|\vey^{(1)},
\vey^{(2)},\vey^{(3)}\ran$ are characterized by the c.m. momentum
$\veP$, and boundstate quantum numbers $n$ and the sum  over $n$
in (\ref{A1.12}) can be written as \be \int\frac{d^3\veP e^{i\veP
\cdot (\veX-\veY)}}{(2\pi)^32\bar \mu_g(n)}
\varphi^+_n(\vexi^{(1)}, \vexi^{(2)}) \varphi_n(\veta^{(1)},
\veta^{(2)}) \label{A1.13} \ee where $\vexi^{(i)};\veta^{(i)}$ are
Jacobi coordinates, while $\veX, \veY$ are c.m. coordinates.

\begin{figure}[!t] 
\epsfxsize=12cm 
\centering
\epsfbox{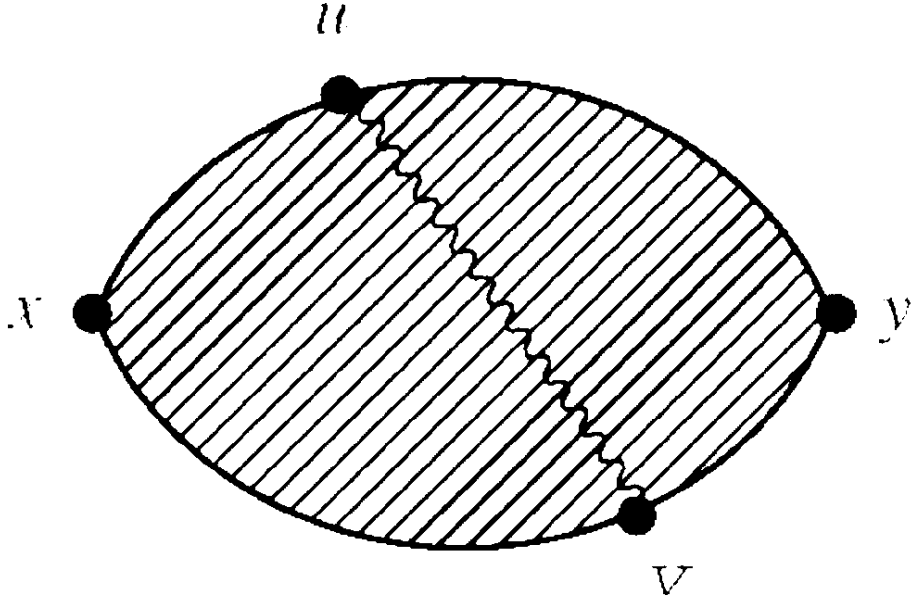} 
\caption{The same as in Fig.1 but with one perturbative gluon
    passing between points $u$ and $v$ described by a wavy line.}
\end{figure}

We are now in the position to write the amplitude, corresponding
to the Feynman diagram of Fig.2, where the valence gluon $a_\mu$
is emitted at the point $u$ and absorbed at the point $v$. The
corresponding term in the Lagrangian is given in (\ref{14}),
 \be
S_1(a,q,\bar q)=\int\bar q(x) g\hat a(x) q (x) d^4 x.
\label{A1.14} \ee

The general form of the amplitude of Fig.2 is
$$
G_{q\bar q}^{(2)} =\int \lan {\rm tr}_{C\alpha} \Gamma^{({\rm
out})} S(x,u) t^a g\gamma_\mu d^4 u S(u,y)  G^{ab}_{\mu\nu} (u,v)
\Gamma^{({\rm in})}\times
$$
\be \times S(y,v) g\gamma_\nu t^bd^4v S(v,x)\ran_B. \label{A1.15}
\ee Here trace is over both color $(c)$ and Lorentz $(\alpha)$
indices.

Introducing now the representations as in (\ref{A1.12}),
 (\ref{A1.13}),  one obtains finally
  $$
  G_{q\bar q}^{(2)} = \frac{N^2_c {\rm tr}_\alpha}{2} \lan \Lambda_q
\Gamma^{(out)} \Lambda_{\bar q} \varphi^+_M(0) \int
\sum_n\frac{\gamma_\mu
   V_{Mn}^{(\mu)} V_{nM}^{(\nu)} \gamma_\nu \varphi_M(0) \Lambda_q
   \Gamma^{(in)}\Lambda_{\bar q}\ran}{2\mu_g(n) (M_h^{(n)}
   -M_M)(M_M-M_h^{(n)})}
   \times
   $$
   \be
   \times \frac{ d^3\veP}{(2\pi)^3} e^{i\veP\cdot (\vex-\vey) -
   E_M(y_4-x_4)}.
   \label{A1.15}
   \ee
   Here notations are used
   \be
   V_{Mn}^{(\mu)}= g\int\varphi_M(\ver)^\mu\psi^+_n(0,\ver) d^3\ver
   \label{A1.15a}
   \ee
   and $\varphi_M(\ver)$ , $^\mu\psi_n$ are meson and hybrid
    eigenfunctions respectively (with the gluon in the latter having
    polarization $\mu$), while $M_M, M_h^{(n)}$ are masses of the
    meson and hybrid (in the  $n$-th excited state) respectively.
    Note that while $G_{q\bar q} $  (\ref{A1.3}) is $O(N_c)$ due to
    ${\rm tr}_c$, $G_{q\bar q}^{(2)}$ is $O(g^2 N^2_c)=O(N_c)$, since gluon
    line in Fig.2 is equivalent to double fundamental line, yielding
    two color traces in Fig.2.

          From (\ref{A1.15}) one can see that the basic element
          defining the amplitude of the mixing of meson and hybrid is
          the dimensionless ratio
          \be
          \frac{V_{on}^{(\mu)}}{\sqrt{2\mu_g(n)}|M_h^{(n)}-M_M|}\equiv \lambda_n^{MH}
          \label{A1.16}
          \ee
          The rest of this section is devoted to the calculation of
          the matrix element $V_{Mn}^{(\mu)}$ using realistic wave
          functions for the meson and hybrid.

          To make estimates of $V_{Mn}$, we first remark that to the
          lowest approximation in spin splittings both $\varphi_M$
          and $^\mu\psi_n$ are proportional to unit  matrices in
          Lorentz indices, and moreover the gluon Green's function
          $G^{ab}_{\mu\nu}$ is proportional to $\delta_{\mu\nu}$,
          where each bound gluon acquires its mass due to the
          attached string (similarly to the $W^\pm, Z^0$ acquiring
          mass due to attached Higgs condensate) and hence the sum
          over $\mu\nu$ is the sum over 3 polarizations.

          As a result $V_{Mn}^{(\mu)}V_{nM}^{(\nu)}\sim\delta_{\mu\nu}$
           and $^\mu\psi^+_n(0,\ver)$ does not depend on
          $\mu$ in the same lowest approximation (when gluon spin
          contribution to the hybrid mass is neglected; both for
          mesons and hybrids the spin splitting is less or about 10\%
          of the total energy and our estimates will have this
          accuracy).

The meson wave function $\varphi_M (\ver)$ for linear growing
potential is Airy function, for our purposes it is enough to use a
Gaussian with the proper radius  $r_0$, e.g. \be
\varphi_M(r)=\left (\frac{3}{2 r^2_0\pi}\right)^{3/4}
e^{-\frac34\left(\frac{r}{r_0}\right)^2}. \label{A1.17} \ee Using
$r_0$ as a variational parameter, one gets \be \lan
r^2\ran=r^2_0=0.725 {\rm fm}^2. \label{A1.18} \ee
 For the hybrid wave
function one can use the eigenfunction of the Hamiltonian
(\ref{10}) in the lowest hyperspherical approximation, where for
an estimate we approximate $W(\rho)$ by the oscillator well
$C(\rho-\rho_0)^2, C\equiv \frac{m\omega^2}{2}$ near its minimal
value (the accuracy of this procedure is around one percent for
the cases considered, see \cite{24})).

Then one can write \be \psi(\ver_{13}, \ver_{32}
)=\frac{1}{\rho^{5/2}\sqrt{\Omega_6}} \left| \det\left (
\begin{array}{lll}
\ver_1 &\ver_2 &\ver_3\\
\veR &\vexi& \veta
\end{array}
\right)\right|^{1/2}
 \left(\frac{m\omega}{\pi}\right)^{1/4}
e^{-\frac{m\omega}{2} (\rho-\rho_0)^2}.
\label{1.18a}
\ee

Here $m$ is an arbitrary mass, used for dimensional reasons and
disappearing from final answers, $\rho_0$ is found from the
minimum of $W(\rho)$, $W'(\rho=\rho_0)=0$, and $w$ is expressed
through $W^{''}(\rho_0), m\omega^2=W^{''}(\rho_0)$, so that \be
\omega=(1.358)^{2/3} \sqrt{3} \sigma^{2/3}
\left(\frac{\mu_1+\mu_3}{\mu_1\mu_3}\right)^{1/3} \frac{1}{[{\cal
L}({\cal{L}}+1)]^{1/6}},{\cal L}=K+\frac32, \label{A1.19} \ee \be
\rho_0=\frac{2}{\sqrt{m}}\left(\frac{{\cal L}({\cal L}+1)}{1.358
\sigma}\right)^{1/3}
\left(\frac{\mu_1\mu_3}{\mu_1+\mu_3}\right)^{1/6}, \label{A1.20}
\ee \be \left| \det\left (
\begin{array}{lll}
\ver_1 &\ver_2 &\ver_3\\
\veR &\vexi& \veta
\end{array}
\right)\right|^{1/2}= \left(\frac{\mu_1\mu_2\mu_3}{\mu
m^2}\right)^{3/4}, ~~~ \mu= \mu_1+\mu_2+\mu_3. \label{A1.21} \ee

The determinant (\ref{A1.21}) appears due to the change of
variables, as follows:
 \be \int d\ver_1 d\ver_2 d\ver_3|\Psi|^2=
\int d\veR d\vexi d\veta |\psi|^2|\det()|= \int d\veR\rho^5 d\rho
d\Omega_6|\Psi|^2|\det()|. \label{A1.22} \ee

Finally $\int d \Omega_6\equiv\Omega_6=\pi^3$ and we need
expression of $\rho$ through $\ver_{13}, \ver_{23}$ etc., namely
\be \rho^2=\veta^2 + \vexi^2 =\sum_{i<j} \frac{\mu_i\mu_j}{m\mu}
(\ver_i-\ver_j)^2. \label{A1.23} \ee

In $V_{Mn}$ enters $\Psi_n$ at $\ver_{13} =0,
\ver_{23}=\ver_{21}\equiv \ver,$ which yields \be
\rho^2=\ver^2\frac{\mu_1(\mu_3+\mu)}{2m\mu}. \label{A1.24} \ee

Insertion  of (\ref{A1.17})--(\ref{A1.21}) in (\ref{A1.15}) yields
\be \frac{V_{Mn}}{\sqrt{2\mu_g}}=Cg\frac{r_0^{3/2} \sigma\mu_1
J}{\sqrt{2\mu_g}}
 \label{A1.25}
  \ee
  where following notations are used:
   \be
J= \int^\infty_0 t^2 dt e^{-t^2-a(t-b)^2}, \label{A1.26} \ee \be
C=\frac{1.358\cdot 2^{17/4}}{\pi^4({\cal L}({\cal L}+1))^{7/8}
3^{5/8}} \left( \frac{\mu_1+\mu_3}{\mu_3} \right)^{1/2}
\left(\frac{\mu_3}{\mu}\right)^{3/4}, \label{A1.27} \ee \be
a=\frac{2\omega r_0^2}{3} \frac{\mu_1(\mu_1+\mu_3)}{\mu},~~
b=\rho_0\sqrt{\frac{3\mu m}{4r^2_0\mu_1(\mu_1+\mu_3)}}.
\label{A1.28} \ee

In (\ref{A1.25}) $\mu_g$ is the gluon effective energy, where gluon
belongs to the hybrid, so that $\mu_g=\mu_3$.

We can do now estimates of $V_{Mon}$ both for light and heavy
quarks. Using \cite{20,1} for massless quarks and for
$\sigma=0.18$ GeV$^2$ one has $ \mu_2=\mu_1=0.312$ GeV,
$\mu_3=0.442$ GeV.\, $\omega=0.956$ GeV, $a=2.54,~~b=0.813,$ and
hence \be \frac{V_{Mn}}{\sqrt{2\mu_3}}= g\cdot 0.2 J\la g\cdot
0.08~ {\rm GeV}. \label{A1.29} \ee For heavy quarks, e.g., for
$m_1=m_2=4.8$ GeV, one has $r_0=0.356$ fm, $\mu_1\simeq m_1=4.8$
GeV, $\mu_3=0.767$ GeV, and $\omega= 0.5$ GeV. Inserting all these
values in (\ref{A1.25}) one obtains again for heavy meson-heavy
hybrid mixing \be \frac{V_{Mn}}{\sqrt{2\mu_3}}= g\cdot 0.08~ {\rm
GeV}. \label{A1.30} \ee This estimate is  close to the
calculations made in \cite{5,6}. Hence one obtains a small mixing
parameter $\lambda_n$ in (\ref{A1.16}) when mass difference
between meson and hybrid is large, $\Delta \sim 1$ GeV, namely \be
\lambda_n\la 0.1 g \label{A1.31} \ee while for close values of
meson and hybrid masses, $\Delta M\sim 0.1$ GeV, the mixing can be
large and one should use the many-channel Hamiltonian,  which can
be obtained from (\ref{10}).

   \section{
   Mixing of glueball with hybrid  and mesons}

\begin{figure}[!t] 
\epsfxsize=12cm 
\centering
\epsfbox{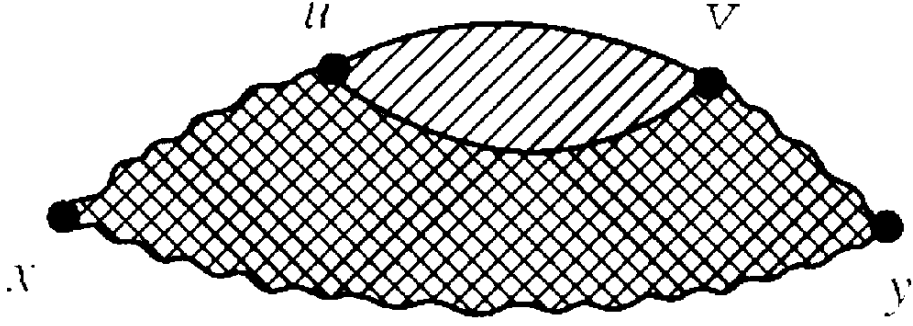} 
\caption{Glueball Green's function describing propagation of two
    gluons (wavy lines) from point $x$ to point $y$, with insertion
    of quark-antiquark propagators (solid lines) from point $u$ to
    point $v$. The  cross-hatched region refers to the
    adjoint string world sheet, while hatched region
    between quark lines has the same meaning as in Fig. 1.}
\end{figure}

Here we consider a two-gluon glueball state, and using the same
methods as discussed in Section 4 and Sections 1,2  we write the
amplitude corresponding to the diagram of Fig.3:

$$
G_{gg}^{(2)\mu\nu,\mu'\nu'} (x,y) \equiv g^2 {\rm tr}_c\{
\Gamma^{({\rm out})} G_{\mu\rho} (x,u)  t^a {\rm
tr}_\alpha(\gamma_\rho S(u,v) \gamma_\sigma S(v,u))\times
  $$
  \be
  \times
  t^b G_{\sigma\nu} (v,y) \Gamma^{({\rm in})} G_{\mu'\nu'} (x,y)\}.
  \label{A2.1}
  \ee
  Representing each gluon line at large $N_c$ as a double fundamental
  line, one gets two closed color contours in Fig.3 and  as a result the
  factor $N_c^2$, hence the amplitude of Fig. 3 is $O(g^2N_c^2)\sim
  O(N_c)$ as well as the leading two-gluon diagram $G_{gg}^{(0)}$
  which means that mixing here also appears in the leading order of
  the $1/N_c$ expansion.

  Using expressions for the Green's functions (\ref{A1.7}),
  (\ref{A1.8}), one obtains that insertion of the quark loop in Fig.
  3 contributes to the glueball Green's function the term
  \be
  G_{gg}^{(2)\mu\nu,\mu'\nu'} =\frac{N_c^2}{2}
{\rm   tr}_\alpha\{\Gamma^{({\rm out})}\varphi^{+(\mu\nu)}_G(0)
  \sum_n\frac{\gamma_\rho \Lambda_q\tilde V_{Mn}^{(\rho)}\tilde
  V_{nM}^{(\sigma)}\Lambda_{\bar q} \gamma_\sigma
  \varphi_G^{\mu'\nu'}(0)}{2\mu_g(M_H^{(n)}
  -M_G)(M_G-M_H^{(n)})}\Gamma^{({\rm in})}\}.
  \label{A2.2}
  \ee
  Here $\varphi _G^{(\mu\nu)}\mu_g$  is the effective
 (constituent) mass of the gluon in the initial
and final glueball (which is calculated
 through $\sigma_{\rm adj}$), $M_G$ is the glueball mass, and other
 notations are the same as in the previous section.

Similarly to the case of
 meson-hybrid mixing we introduce the mixing parameter
\be \frac{\tilde V_{on}}{\sqrt{2\mu_g(G)}|M_H^{(n)}-M_G|}=
\lambda_n^{GH}, \label{A2.3} \ee
 where $\tilde V_{Mn}$ is the same as
in meson-hybrid case, (\ref{A1.15}) with the replacement of meson
to the glueball Green's function \be \tilde V_{Mn}
 =g\int\varphi_G(\ver)
 \psi^+_n(\ver_{12}=0, \ver_{13}=\ver_{23}=\ver) d^3\ver.
\label{A2.4} \ee To estimate $\tilde V_{Mn}$ one can use the
 same  (\ref{A1.25})  with following replacements:

(i)$\mu_g$ is the effective mass
 (energy) of the gluon in the glueball, for lowest glueball states
$\mu_g=0.53$ GeV $(L=n_r=0), 0.69$ GeV $(L=1, n_r=0)$ \cite{16,1};

(ii) $ a\to \tilde a =
 \frac{2\omega r^2_0}{3}
 \frac{\mu_3(\mu_1+\mu_2)}{\mu}, ~~  b\to \tilde b=\rho_0\sqrt{\frac{
3\mu m}{4r^2_0\mu_3(\mu_1+\mu_2)}}$.

Insertion of the values of $\mu_1,\mu_2, \mu_3,\omega, r_0$ into $\tilde a,
\tilde b$ yields $\tilde a=1.17
 a, ~~\tilde  b=0.923 b$, hence
 all estimates for $V_{Mn}$ made in (\ref{A1.29}),
(\ref{A1.30}) hold also for $\tilde V_{Mn}$, and the
glueball-hybrid mixing
 is the same as meson-hybrid one,
 provided the radius $r_0$ of glueball is  the same as that of meson.

\begin{figure}[!t] 
\epsfxsize=12cm 
\centering
\epsfbox{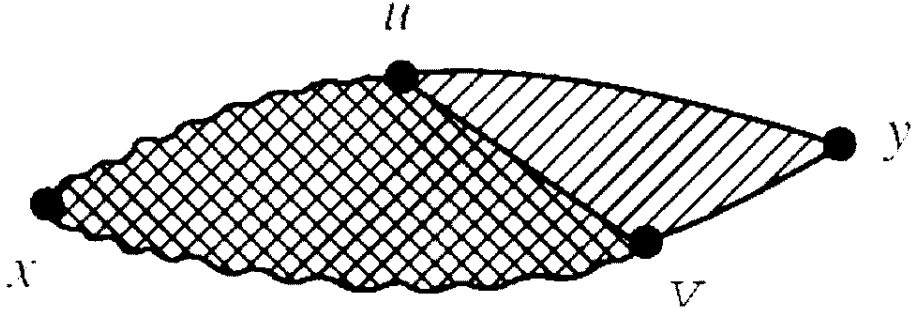} 
\caption{The amplitude (Green's function) for transition of
a    glueball into a meson.  Symbolic notations are the same as in
    Figs.1 and 3.}
\end{figure}

We come  now to the amplitude of Fig.4,
 which gives the amplitude for
 glueball-meson mixing. Analysis similar to the previous one
 yields for this amplitude the following ratio:
\be \lambda_n^{GM} =\frac{\tilde V_{Mn}}{\sqrt{2\mu_g(G)}}
\frac{V_{nM}}{\sqrt{2\mu_g(H)}}
\frac{1}{|M_G-M_H^{(n)}||M_H^{(n)}-M_m|}= \lambda_n^{GH}
\lambda_n^{HM}. \label{A2.5} \ee Insertion of our results
(\ref{A1.29})--(\ref{A1.30})
 leads to the following order of magnitude estimate
\be \lambda_n^{GM} \approx N_c\frac{g^2(0.08~{\rm GeV})^2}{(\Delta
M)^2}. \label{A2.6} \ee

For $\Delta M\sim 1$ GeV one obtains
 $\lambda_n^{GM}\sim1/16$, and the probability of admixture of the
 glueball to a meson $P_{GM}
= |\lambda_n^{GM}|^2$
is less than a percent.
 This should be compared to the
 probability of the hybrid in the same meson
 \be
 P_{HM}
=|\lambda_n^{HM}|^2= N_c g^2 \left( \frac{0.08~{\rm GeV}}{\Delta
M}\right)^2 \approx \lambda^{GM}_n\sim 5-10\%. \label{A2.7} \ee

All these estimates hold for unsuppressed
 glueball-hybrid-meson transitions, when spin-flip
  amplitudes do not enter, otherwise one
may expect additional one order-of-magnitude suppression of
probabilities.

\section{Discussion and conclusions}

In Sections 4 and 5 we have obtained reduced matrix elements for
meson-hybrid, hybrid-glueball, and meson-glueball transitions. The
full matrix elements involve expressions of the following type
(see  (\ref{A1.14})
 and (\ref{A2.2})):
 \be
 K^{MH}_J\equiv
 \int(\Phi^{(M)}_J)^+_{\alpha\beta}\gamma_\mu\Phi^{(H)}_{J,\mu}
 d\tau_3,
 \label{74}
 \ee
 where
$\Phi^{(H)}_{J,\mu}$ contains all components of the hybrid wave
function.

 Since the Hamiltonian (\ref{10}) and the eibein technic
 (\ref{11}) separate only positive values of $\mu_q,\mu_{\bar
 q}$, and $\mu_g$, the wave functions $ \Phi^{(M)}_J$ and
 $\Phi^{(H)}_J$ refer to those  positive values and we disregard
 the negative energy components (negative $\mu$-components).

 Moreover, spin interactions in mesons, hybrids, and glueballs are
 treated perturbatively in our formalism, and comparison to
 lattice data for the same input shows that both approximations of
 positive $\mu_i$ and perturbative spin splittings are rather good
 for the cases considered -- see \cite{1} for Tables with
 comparison. These approximations are not good for pion and kaon,
 where both spin and negative energy states play an important
 role, but we shall not consider these mesons, or hybrids with
 similar properties.

 Therefore our parameters  (\ref{A1.16}) and (\ref{A2.3}) refer only
 to such states which are well described by the Hamiltonian
 (\ref{10}) with condition (\ref{11}). In this case the matrix
 element of $\gamma_\mu$ is simplified since the $\mu$ dependence
 in the hybrid
 decouples from the spin (Lorentz) indices and the simplifications
 made after (\ref{A1.16}) are valid, when one neglects possible
 Clebsch--Gordon coefficients, which are of the order of unity.

 In this way one obtains an estimate for the ``superallowed"
 transitions meson-hybrid-glueball, not involving spi-flip or
 negative-sate components.

 A comparison with the results of \cite{5}, \cite{6} shows a good
 agreement within a factor of two, which tells that the missing in
 our case factors are of the order of unity. However, in
 \cite{5},\cite{6} as well as in our case only wave functions were
 considered without negative energy components (negative $\rho$-spins).

 To have more comparisons, one may look at the lattice calculation
 of MME, which have been done extensively for the glueball
 $0^{++}$- scalar meson case \cite{25,26}.

 The authors \cite{26} arrive at the following result
 of careful lattice studies:\\

 $$|f_0(1710)>=0.859|g>+0.302|s\bar s>+0.413|n\bar n>, $$

 $$|f_0(1500)>=-0.128|g>+0.908|s\bar s>-0.399|n\bar n>,
 $$
\be
 |f_0(1390)>=-0.495|g>+0.290|s\bar s>+0.819|n\bar n>.
\label{100} \ee

>From (\ref{100}) it is clear that a strong mixing occurs between
states in the region 1.4--1.7 GeV. However our results for MME
between glueball and meson  states always implied an appearance of
a hybrid state as an intermediate  state between glueball and
meson. From (\ref{A2.5})
 it is clear that the strong glueball-meson
 mixing is possible only if the
 intermediate hybrid has the mass in the same interval
  1.4--1.7 GeV  introducing in (\ref{A2.6}) $\Delta
  M \sim 0.2-0.3 $ GeV, one obtains $\lambda_n^{GM}\approx 1,$
  as in the lattice calculations \cite{26} for the $0^{++}$ states.

  Hence one could look for the
  $0^{++}$ hybrid state in the discussed mass range.
   Indeed, analytic calculations of lowest hybrid states in \cite{27}
  confirm the possibility of the $0^{++}$
state around 1.7-1.8 GeV. this problem clearly calls for further
theoretical and experimental investigation.

The author is grateful to K.A.Ter-Martirosyan for many stimulating discussions, suggestions and criticism of the first drafts of the paper.
This work was partially  supported by the RFFI grants 00-02-17836 and 00-15-96786.

 \setcounter{equation}{0}
\renewcommand{\theequation}{A.\arabic{equation}}
\begin{center}

{\bf Appendix\\
Meson wave-functions with definite $J^P$}\\
\end{center}

In (\ref{21}) the wave function of the meson was given in the
so-called $4\times 4$ representation, where the total angular
momentum $J$ and parity $P$ were not specified.

In this Appendix some relations and notations are given for meson
wave functions with given $J^P$, based on last reference
\cite{23}.

The states of the quark may be classified according to the
so-called $\rho$-spin and usual spin states, where $\rho$-spin
states, $\rho=\pm 1$, can be taken as eigenvalues of each of
$\gamma_0^{(i)}~(i=1,2$ refer to quark 1 and quark 2, which may be
an antiquark), or as eigenvalues of $\Lambda^\pm$, where

 \be
\Lambda^\rho(\vep)
=\frac{\rho(\vegam\vep+m)+\gamma_0\sqrt{\vep^2+m^2}}{2\omega}.
\label{a.1} \ee

In both cases one can create four $\rho$-spin eigenstates (e.g.,
as eigenstates of $\gamma_0^{(1)}\gamma_0^{(2)}$):

$$
s=\frac{|++>-|-->}{\sqrt{2}}, ~~ a=\frac{|++>+|-->}{\sqrt{2}},
$$
\be e=\frac{|+->+|-+>}{\sqrt{2}},~~ O=\frac{|+->-|-+>}{\sqrt{2}}.
\label{a.2} \ee

Using \cite{22} one can form  8 states of unnatural parity  for
$qq, P=(-)^{J+1}$ \be
B=\{~^1J^s_J,~^1J^a_J,~^3J^s_J,~^3J^a_J,~^3(J-1)^e_J,~^3(J-1)^0_J,~^3(J+1)^e_J,
~^3(J+1)^0_J\}, \label{a.3}\ee and 8 states of natural parity
$P=(-)^J$:

\be
B^*=\{~^1J^e_J,~^1J^0_J,~^3J^e_J,~^3J^0_J,~^3(J-1)^s_J,~^3(J-1)^a_J,~^3(J+1)^s_J,
~^3(J+1)^a_J\}. \label{a.4}\ee

Dynamical equations for mesons connect all components belonging to
the same $J^P$.

The 16, component wave function (\ref{21}),
$f^{G_1,G_2}_{\alpha_1\alpha_2}$, can be decomposed into $4\times
4$ $q\bar q$ basis, introduced in the last reference \cite{22}, as
it is shown in the Table below.

\begin{table}
\caption{ 16-component vector $qq$-basis states and corresponding
$4\times 4$ matrix $q\bar q$-basis states. (The angular dependence
of the wave functions of the $q\bar q$-triplet states must
determine to which $qq$-triplet state they correspond. For a
correct normalization, all matrix  states should be multiplied by
1/2; furthermore, matrix states proportional to $p$ or $p$ need an
extra factor $1/M $ and $1|p|$, respectively. The correspondence
is only valid in the c.m.s. and $p_0=0$ is assumed.) Here $P$ and
$p$ refer to the total and relative momentum of $q\bar q$ in the
meson respectively.}

\bigskip
\bigskip
 \begin{tabular}{|l|l|}\hline\hline
 $qq$-State&$q\bar q$-State\\
$~^1J^s_J$&
 $-\gamma_5\not{P}$\\
 $~^1J^a_J$&
 $-\gamma_5$\\
 $~^1J^e_J$&-1\\
 $~^1J^0_J$&
$ \not{P}$\\
$~^3J^s_J,~^3(J-1)^s_J,~^3(J+1)^s_J $&$-\not p$\\
$~^3J^a_J,~^3(J-1)^a_J,~^3(J+1)^a_J$ &$i\sigma_{\mu\nu} P^\mu
p^\nu$\\
$~^3J^e_J,~^3(J-1)^e_J,~^3(J+1)^e_J$ &$-\gamma_5i\sigma_{\mu\nu}
P^\mu
p^\nu$\\
$~^3J^0_J,~^3(J-1)^0_J,~^3(J+1)^0_J$ &$\gamma_5\not p$\\
\hline \hline
  \end{tabular}
 \end{table}

 A similar expansion can be derived for the hybrid states, which include in general $16\times 3$
 components. In the estimates of the present paper it is assumed
 that the positive energy (positive $\rho$-spin) component gives
 the dominant contribution to the mixing matrix element, and that
 spin interaction can be treated perturbatively. Then the
 estimates (\ref{A1.16}), (\ref{A2.3}) are  modified due to kinematical
 coefficients (like those in the Table) by factors of the order of
 unity.

        \end{document}